\documentclass[11pt,preprintnumbers]{article}
\usepackage{moriond,epsfig}

\def\Journal#1#2#3#4{{#1} {\bf #2}, #3 (#4)}

\def\PRD{{\em Phys. Rev.} D}

\def\be{\begin{equation}}
\def\ee{\end{equation}}
\def\bea{\begin{eqnarray}}
\def\eea{\end{eqnarray}}

\begin{document}

\begin{flushright}
LAPTH-1116/05
\end{flushright}

\vspace*{3.5cm}
\title{Ultra High Energy Cosmic Rays from charged black holes.\\
A new theoretical possibility?}

\author{ ALVISE MATTEI }

\address{LAPTh -- Laboratoire d'Annecy-le-Vieux de Physique Th\'eorique, 9 chemin de Bellevue,\\
Annecy-le-Vieux BP110, F-74941, France}

\maketitle\abstracts{In General Relativity, there is a new field of activity concerning the study of charged stars. In a recent paper, Ray et al. have shown the possibility that the collapse of a charged star could form a charged black hole before all the charge leaves the system. In this field of view we propose a new model for UHECR and we will show that it is possible to accelerate cosmic rays up to EeV. In this talk we will compute the UHECR flux, the charged black hole density and the energy spectrum associated with them in order to reproduce experimental data. We will see that we need only a small number of these hidden sources in order to account for observed UHECR properties and we  will  study the  limits on charge and mass of  black holes.}

\section{The charge on astrophysical objects}
As it was known since 1924, every self-gravitating system has a net charge. This can be obtained moving from the classical Poisson's equations \cite{r24}.
For \emph{isothermal equilibrium}, densities can be expressed in terms of Boltzmann distributions, which yield, in terms of expected values for energy~\cite{a80}
$-e\psi\simeq Am_p \phi+Ze\psi\Longrightarrow e\psi\simeq -{A}m_p\phi/{(Z+1)}$, or in terms of charge to mass density ratios ${\rho_e}/{\rho}= {GAm_p}/{(Z+1)e}$.
For a classical gravitational potential $\phi=GM/R$ and electric potential $\psi=Q/R$. So the net charge of a star of mass $M$ is about 150~C.

The {\it potential differential} between the center and the edge of a self-gravitating system is \cite{bh78}
\begin{equation}
V=\frac{A}{Z+1}\frac{m_p}{e}\frac{GM}{R}\simeq \frac{A}{(Z+1)}\frac{M}{M_\odot}\frac{R_\odot}{R} \times 1900~\mathrm{Volts}
\end{equation}
For a {\it neutron star} this would be $V\sim 10^8$~Volts. We notice that a self-gravitating object built with particles of extremely high A/Z ratio would have a huge charge.

\section{General Relativity Limits: compact stars and black holes}
Once understood that any self-gravitating object has a charge, here we want to evaluate the upper limits on charge that arise from general relativity.

In radial coordinates, we can write the metric of a spherically symmetric ball of charged perfect fluid $ds^2=-e^{\nu(r,t)} dt^2+e^{\lambda(r,t)} dr^2+r^2(d\theta^2+\sin^2\theta d\phi^2)$. The stress energy tensor is
$T^{\mu\nu}=(\rho+P)u^\mu u^\nu + Pg^{\mu\nu}+\frac{1}{4}\pi\left[F^{\mu\alpha}F^\nu_{\phantom{\nu}\alpha} - \frac{1}{4}g^{\mu\nu} F^{\alpha\beta}F_{\alpha\beta} \right]$, with $u^{\mu\nu}$ 4-velocity, $g^{\mu\nu}$ metric tensor, $\rho$ mass density, $P$ pressure and $F^{\mu\nu}$ electro-magnetic tensor.
Solving the Einstein equation together with the spherical symmetry condition and the Maxwell equations, one obtains the hydrostatic equations for a compact charged star, and notably a generalized Oppenheimer-Volkoff equation \cite{b71}
\begin{equation}
-\frac{\partial P}{\partial r}=\underbrace{-\frac{q}{4\pi r^4}\frac{\partial q}{\partial r}}_{Coulomb~term}+\underbrace{(\rho+P) \frac{4\pi r^4 P+mr-q^2}{r(r^2-2mr+q^2)}}_{Regeneration~term}
\end{equation}
where $q(r,t)=  \int_{0}^{r} e^{(\lambda+\nu)/2} 4\pi r^2 j^0 dr$ is the charge inside a shell, and $j^0$ is the zero component of current density $j^\mu$. The charge is thus playing a double role, because it contrasts the gravitational attraction as well as it contributes to the increase in gravitational mass. So it arises a constraint on upper limits on charge.

The complete structure of a spherically symmetric, charged, compact star was calculated for a polytropic equation of state and a charge proportional to density \cite{remlz03}. 
It gives $Q\simeq10^{20} {M}/{M_\odot}${~C}.
The same equation was solved for a \emph{stable quark star}, with an EOS $\rho=3P+4B$, that is the so called MIT bag model, and it was found a huge surface field \cite{afo86,kwwg95} $\mathcal{E}=10^{19}$~V/m.

Other constraints can be fixed on black holes.
A charged static black hole is given by Reissner-Nordstr\"om metric, that defines the horizon radius as $r_\pm=M\pm\sqrt{M^2-Q^2}$
To avoid naked singularity, we need to stay below
\begin{equation}
Q_{M}=\sqrt{4 \pi \epsilon_0 G}M\simeq 1.7\times 10^{20} \frac{M}{M_\odot}~\mathrm{C}
\end{equation}
A large electric field can be induced by an external magnetic field \cite{h82}
$B\sim E$
\begin{equation}
Q\simeq1.2\times10^{14}\left(\frac{M}{M_\odot}\right)^{3/2}\left( \frac{B}{10^{13}\mathrm{G}}\right)^{1/2} \mathrm{C}.
\end{equation}

A maximally rotating black hole, described by a Kerr metric, in a magnetic field aligned along its symmetry axis will accrete charge until it becomes \cite{w74}
\begin{equation}
Q=2B_0 J \times \left(\frac{4\pi \epsilon_0 G}{c^2} \right)\leq1.5\times 10^{14}\left(\frac{M}{M_\odot}\right)^2\left(\frac{B}{10^{13}~\mathrm{G}} \right)~\mathrm{C}.
\end{equation}
Charge was treated as a perturbation of Kerr metric and cannot force the Reissner-Nordstr\"om limit.
Other calculations were made on this, with different magnetosphere structure, finding similar values of charge \cite{rl}.\\
{\bf QED limits:} Pair creation puts a strong limit on electric field at surface of compact stars. The critical field is \cite{s51} $\mathcal{E}_c={m_e^2c^3}/{\hbar e}=1.2\times10^{18}$~V/m and the relative magnetic field limit is $B_c=4\times 10^{13}$~G.
If $\xi=Q/Q_{M}$, for a black hole the upper limit on electric field is
\begin{equation}
\mathcal{E}(r_+)<\mathcal{E}_c \Longrightarrow \frac{M}{M_\odot}>6\cdot 10^5 \frac{\xi}{(1+\sqrt{1-\xi^2
})^2}
\end{equation}
{\small
\begin{table}[h]
\caption{\small Potential differential at surface for different stellar objects. Maxima are corrected for QED limits}
\begin{center}
\newpage
\begin{tabular}{c c c}
\textsc{Type} & \textsc{Min} & \textsc{Max}  \\
\hline
\textsc{Normal star} & $1900$~V & $7\times 10^{26}$~V \\
\textsc{Compact star} & $10^8$~V & $2\times 10^{22}$~V \\
\textsc{Black hole} & $10^9$~V & $1.5 \times 10^{21}\frac{M}{M_\odot}$~V  \\
\hline
\end{tabular}
\end{center}
\end{table}%
}

\section{Charged black holes as UHECR accelerators: energy spectrum derivation, time duration and flux}
If an extremely charged black hole exists, for instance soon after a collapse, one of its signature can be the UHECR spectrum. Here we propose a simple model to evaluate the emission of such objects. A charged black hole is surrounded by an Hydrogen cloud, which is the reservoir for particle to be accelerated. The cloud is ionized if the electric field exceeds the ionization field for a H atom: $\mathcal{E}>V_{ion}/a_0$, $V_{ion}=13.6$~V, $a_0=0.5\times 10^{-10}$~m. The acceleration zone is then constrained between the horizon and an outer radius, where the electric field falls below the above limit. The extracted protons are accelerated at energy comprised between the following limits:
\begin{eqnarray}
E_{max}\sim \frac{Q}{4\pi\epsilon_0 r_+}\simeq  5\,\xi\times 10^{26}~\mathrm{eV}&;&
E_{min}\sim \sqrt{\xi\frac{M}{M_\odot}}\, 6.4\times10^{20} ~\mathrm{eV}
\end{eqnarray}

To calculate the spectrum, we made three simple assumptions: the matter surrounding the black hole is the reservoir for UHECR; each remnant atom has the same ionization probability; spherical symmetry of remnant. The particle number contained in a radius $R$ is $N(R)=\int_{r_+}^{R}4\pi r^2 n(r)dr$
where $n$ is the remnant number density. The particle energy gain is proportional to the scaling of potential, $E\propto r^{-1}$.
So the spectrum is
\begin{equation}
\frac{dN}{dE}dE=4\pi n(r)\zeta^3 E^{-4}dE
\end{equation}
where $\zeta$ accounts for the efficiency of this process. We know that a self-gravitating acretion disk has~\cite{b52} $n(r)\propto r^{1/(\gamma-1)}$ 
where $\gamma$ is the usual polytropic index. 
Then we can write for spectrum:
\begin{equation}
\frac{dN}{dE}dE = AE^{\frac{5-4\gamma}{\gamma-1}}dE
\end{equation}
For a monoatomic gas, $\gamma \sim 5/3\Longrightarrow \alpha\sim2.5$

Also the duration of emission can be calculated as a toy model, under the assumptions of ($i$) matter density constant $n_0$ inside radius $r_0$, ($ii$) stationary supplied matter in time $\tau$ and ($iii$) constant mass during discharge.
So charge will decrease of
\begin{equation}
\dot{Q}=-\frac{4\pi}{3}\frac{n_0 r_0^3}{\tau}Q.
\end{equation}
That is $\dot{\xi}=-{\xi^{3/2}}/{T}$, with $Q=\xi Q_{M}$ and
\begin{equation}
T^{-1}=\frac{4\pi}{3}\frac{n_0 r_{ion}^3}{N_0\tau}\left(\frac{M}{M_\odot}\right)^{1/2}=6.5\cdot 10^{-5}\frac{1}{\tau}\left(\frac{M}{M_\odot}\right)^{1/2}~\mathrm{s}^{-1}
\end{equation}
The characteristic dimensions are then $N_0=10^{39}$, the charged~particle~number $r_{ion}= 2.5\cdot10^9$~m, the outer ionization radius $n_0 = 10^6$~m$^3$, that is the usual ISM density. The solution is
\begin{equation}
\xi=\xi_0\left(1+\frac{\xi_0^{1/2}}{2T} t\right)^{-2}.
\end{equation}
So the discharge time is $\Delta t=3\cdot 10^8\tau\left({M}/{M_\odot}\right)^{-1/2}$~s.
We expect $\tau > 10$~s (time requested for electron to fall inside horizon after ionization), so UHECR bursts duration is $\Delta t\sim 1\div 1000$~yr
\begin{figure}
\epsfig{figure=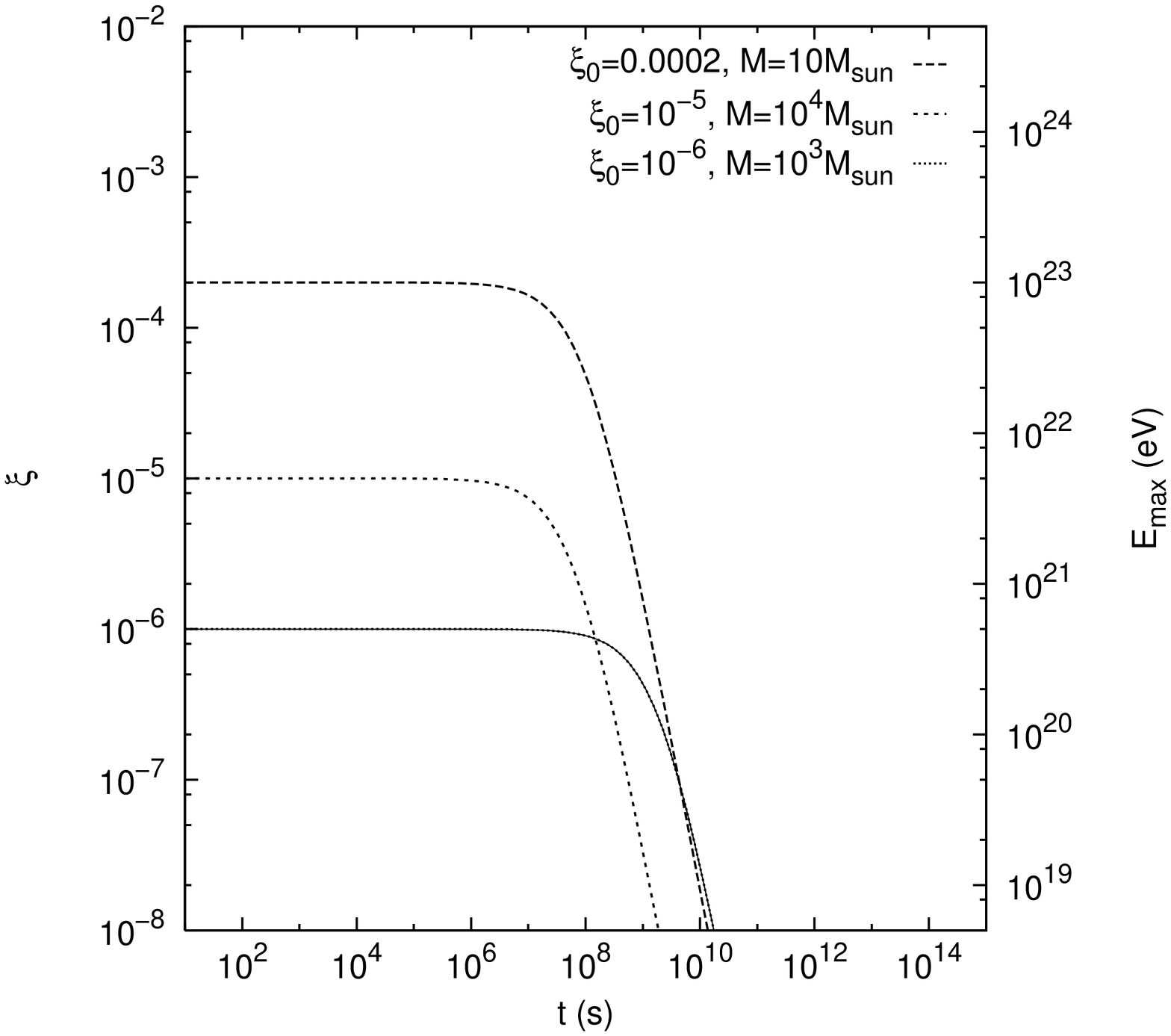,height=2.8in,angle=270}
\hfill
\epsfig{figure=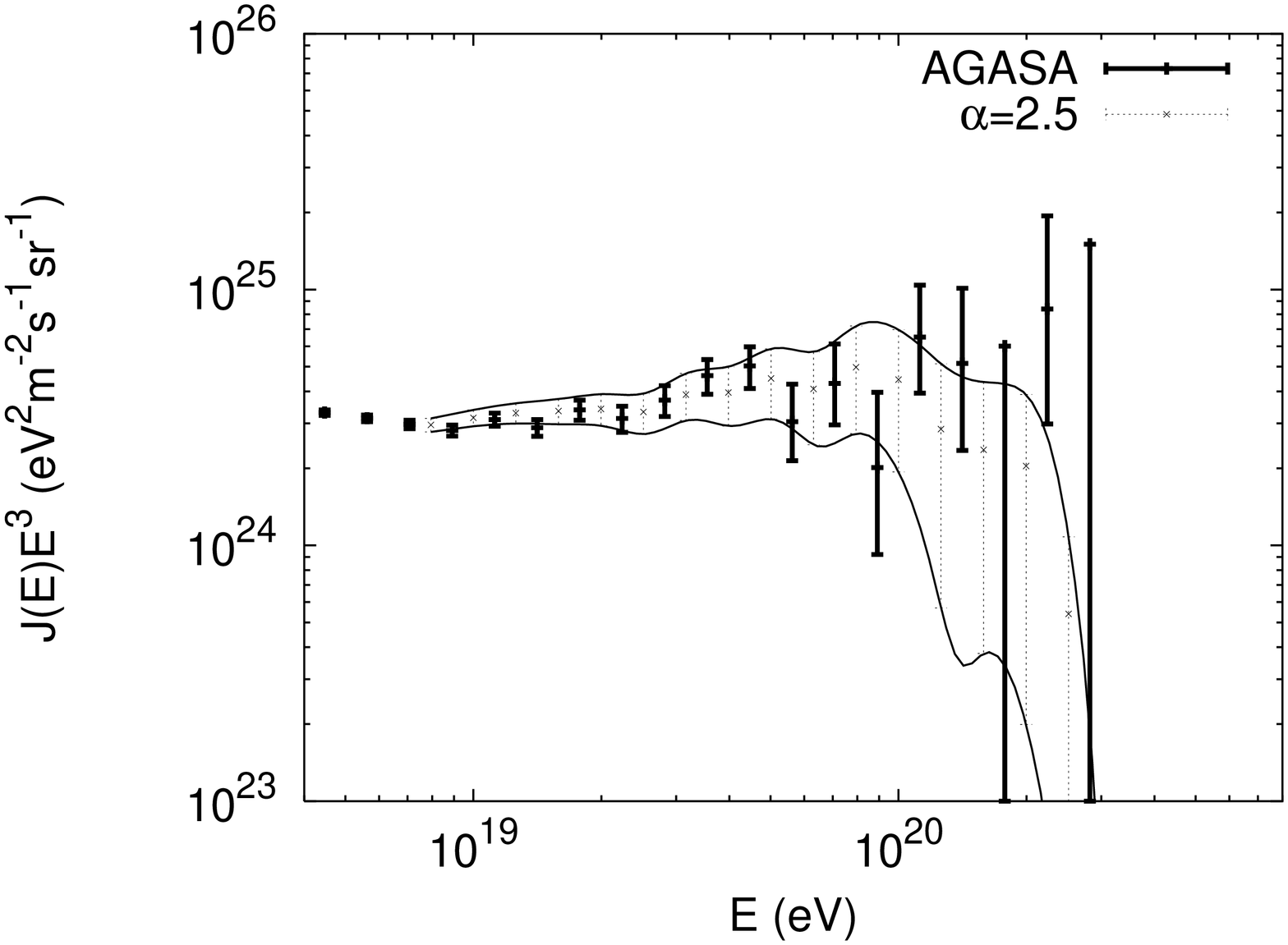,height=3.1in,angle=270}
\caption{Temporal evolution of charge $\xi=Q/Q_M$ in a black hole and spectrum of an homogeneous distribution of such sources, calculated as a MonteCarlo propagation.
\label{fig:chargevol}}
\end{figure}

The flux at $10^{20}$~eV is $\Phi_{obs}=3\cdot 10^{-16} ~\mathrm{m}^{-2}~\mathrm{sr}^{-1}~\mathrm{s}^{-1}$. The flux incoming from a single source at $R_{10}=10$~Mpc is
\begin{equation}
\Phi=\frac{N}{4\pi R^2_{10} \Delta t}=2.8 \times 10^{-17} \frac{M}{M_\odot}\frac{Q}{Q_M}\Delta
t_{yr}^{-1} ~\mathrm{m}^{-2}~\mathrm{sr}^{-1}~\mathrm{s}^{-1}.
\end{equation}
Few black holes
with ${M}/{M_\odot}=10^2$ and ${Q}/{Q_{M}}=10^{-3}$ in a GZK volume can account for present UHECR flux, that is a source density $\sim10^{-5}$~Mpc$^{-3}$ (within 95\% CL of AGASA source density~\cite{fk01}).
Expected active sources may be obtained for instance from local GRB rate ($\sim 10^{-8}$~Mpc$^{-3}$~yr$^{-1}$ for isotropic emission) integrated through lifetime. This result does not depend on the process efficiency $\zeta$, because we exactly know how many particles have to be accelerated during the time $\Delta t$.

As a conclusion, we can say that a charged black hole can accelerate protons up to extremely high energy, with a power spectrum according to present observations and well focused on high energy. The observed multiplets can also be explained through the long emission time. Possible further traces of such electric black holes could be the gamma ray bursts, whose rate is able to recover the observed flux.

\end{document}